\documentclass[letterpaper,twocolumn,english,prl]{revtex4}
\usepackage[OT1]{fontenc}
\usepackage[latin1]{inputenc}
\usepackage{babel}
\usepackage{graphics}

\makeatletter

\providecommand{\LyX}{L\kern-.1667em\lower.25em\hbox{Y}\kern-.125emX\@}

\usepackage{pslatex}

\makeatother
\begin{document}

\title{The Gap at \( \nu =5/2 \) and the Role of Disorder in Fractional
Quantum Hall States}

\author{R. Morf\( ^{(1)} \) and N. d'Ambrumenil\( ^{(2)} \) }

\affiliation{\( ^{(1)} \) Condensed Matter Theory, Paul Scherrer Institute, CH-5232
Villigen, Switzerland\\
\( ^{(2)} \)Department of Physics, The University of Warwick, Coventry,
CV4 7AL, UK}

\date{\today}

\begin{abstract}
Theoretical results for the gaps of fractional quantum Hall states
are substantially larger than experimental values determined from
the activated behaviour of charge transport. The disparity in the
case of the enigmatic \( \nu =5/2 \) state is worrying as it amounts
to a factor 20 to 30. We argue that disorder effects are responsible
for this disparity and show how intrinsic gaps can be extracted from
the measured transport gaps of particle-hole symmetric states within
the same Landau level. We present new theoretical results for gaps
at \( \nu =5/2 \) and 7/2, as well as at \( \nu =1/3, \) 2/5, 3/7
and 4/9, based on exact diagonalizations, taking account of the finite
thickness of the two-dimensional electron layer and Landau level mixing
effects. We find these to be consistent with the intrinsic gaps inferred
from measured transport gaps. While earlier analyses (Du et al, Phys.
Rev. Lett. \textbf{70}, 2944 (1993)) assumed constant broadening for
each sample, our results for the disorder broadening depend on the
filling fraction, and appear to scale with the charge of the elementary
excitations of the corresponding fractional state. This result is
consistent with quasiparticle mediated dissipative transport.
\end{abstract}
\maketitle
The gaps obtained from analyzing the activated temperature-dependence
of the longitudinal conductance near the center of fractional quantized
Hall (FQH) plateaus in GaAs heterostructures \cite{Willett_88,Du93}
disagree with the values obtained from direct diagonalizations of
finite systems \cite{Morf_NdA_SdS_02}. This is the case even taking
account of the softening of the Coulomb interaction between electrons
resulting from the non-zero thickness of the 2D electron layer and
of Landau level mixing effects. The discrepancies can be around a
factor of two in the highest mobility samples for FQH states in the
lowest Landau level (LLL). For FQH states in the second Landau level
the discrepancies are even larger, as much as a factor of 20 at filling
fraction \( \nu =5/2 \) \cite{eisenstein_52_1,pan_exact_quantization,Morf_NdA_SdS_02}
and a factor of 30 at \( \nu =7/2 \) \cite{eisenstein_2002}. Such
large discrepancies make one wonder whether the \( \nu =5/2 \) state
has been correctly identified.

Here, we argue that disorder effects are responsible for these discrepancies.
We show how the intrinsic gap of FQH states, that are strongly affected
by disorder, can be estimated directly from measurements of the transport
gaps using a simple model. These estimates are consistent with results
we obtain from exact diagonalizations of finite systems provided we
take account both of the non-zero width of the electron wavefunction
in the direction perpendicular to the two-dimensional electron layer
and of Landau level mixing (LLM). Our analysis also provides estimates
for the reduction of the measured activation gaps relative to the
disorder-free intrinsic gaps---the so-called `disorder broadening'.
We find that the disorder-induced gap reduction depends on the FQH
state studied and is roughly proportional to the fractional charge
of the corresponding elementary excitations. This is not in agreement
with previous analyses of FQH gaps near \( \nu =1/2 \) which assumed
a filling factor independent disorder broadening of electronic states
\cite{Du93}. These gave intrinsic gaps which scaled as \( |B_{\nu }-B_{1/2}| \),
where \( B_{\nu } \) is the magnetic field at which the filling is
precisely \( \nu  \), and implied an effective mass of the composite
fermions which is independent of the filling fraction, contrary to
theoretical predictions \cite{HLR}. Our results show that, using
a combination of the scaling analysis (described here) and comparisons
with the results of exact diagonalizations, it will be possible to
extract from measurements of activation energies reliable estimates
both for the intrinsic FQH gap and for the disorder-induced gap reduction. 

This work has been motivated by the observation of a transport gap
for a FQH state at \( \nu =7/2 \) by Eisenstein et al. \cite{eisenstein_2002},
and by their report of transport gaps for the \( \nu =7/2 \) and
5/2 FQH states of \( \Delta ^{a}(7/2)=0.07 \)K and \( \Delta ^{a}(5/2)=0.31 \)K.
The latter is almost a factor of three larger than the earlier value
\( \Delta ^{a}(5/2)=0.11 \)K (first reported in reference \cite{eisenstein_52_1}
and confirmed in \cite{pan_exact_quantization}). The smaller gaps
at \( \nu =5/2 \) were obtained for samples with electron density
\( n_{S}=2.3\times 10^{11} \)/cm\( ^{2} \), whereas the most recent
results are for \( n_{S}=3\times 10^{11} \)/cm\( ^{2} \) \cite{eisenstein_2002}.
The factor of three difference between the new and old results for
\( \Delta ^{a}(5/2) \) cannot result from Coulomb interaction effects
alone, as these scale with \( \sqrt{n_{S}} \).

A FQH state at \( \nu =7/2 \) is expected on theoretical grounds.
Its structure should be very similar to that of the \( \nu =5/2 \)
state, as these two states are related by particle-hole conjugation
symmetry, which becomes exact in the limit when LLM can be neglected.
In that limit, if the energy gaps are purely controlled by the Coulomb
interaction,

\begin{equation}
\label{eq:ec}
E_{c}=e^{2}/\kappa \ell _{0},
\end{equation}
 the intrinsic gaps \( \Delta ^{i}(\nu ) \) of pure (disorder-free)
systems can be written as

\begin{equation}
\label{eq:deltai}
\Delta ^{i}(\nu )=\delta (\nu )\, E_{c}.
\end{equation}
The symbol \( \ell _{0} \) in equation (\ref{eq:ec}) stands for
the magnetic length, defined in terms of the magnetic field \( B \),
by \( \ell _{0}=\sqrt{\hbar c/eB} \), and \( \kappa  \) is the dielectric
constant of the semiconductor material. For physically equivalent
FQH states at fillings \( \nu  \) and \( \nu ' \) (those related
by particle-hole conjugation symmetry), the coefficients \( \delta (\nu ) \)
and \( \delta (\nu ') \) in (\ref{eq:deltai}) will be the same and
the difference in the gap values \( \Delta ^{i}(\nu ) \) and \( \Delta ^{i}(\nu ') \)
will reflect the difference in the Coulomb energy scale \( E_{c} \)
at the magnetic fields \( B_{\nu } \) and \( B_{\nu '} \) at which
the FQH states occur. This will happen for \( \nu '=2-\nu  \), and
in the second Landau level, when \( \delta (2+\nu )=\delta (2+(2-\nu )) \),
implying \( \delta (5/2)=\delta (7/2) \). If, in addition, spin-mixing
effects can be neglected, FQH states at filling fraction \( \nu  \)
can be mapped to states at \( \nu '=1-\nu  \). As an example, we
expect that gaps of fractional states at \( \nu =1/3,\, 2/3,\, 4/3 \)
and \( 5/3 \) will all be described by the same coefficient \( \delta (1/3) \)
as long as the Zeeman energy is large enough to suppress spin reversal
in all these states, and as long as LLM effects can be neglected.

On the basis of equation (\ref{eq:deltai}), we analyze the measured
gaps of symmetry related states as a function of \( E_{c} \) to extract
an estimate of the dimensionless coefficient \( \delta (\nu ) \).
In Figure 1, we show the gap results from \cite{eisenstein_2002}
for the \( \nu =5/2 \) and 7/2 states as a function of \( E_{c} \).
The slope of the straight line through the two gap values yields \( \delta (5/2)\approx 0.014 \).
This is just \( \sim 35\% \) smaller than the theoretical estimate
for a spin-polarized paired state of the Moore-Read type \cite{Moore-Read},
\( \delta ^{th}(5/2)\approx 0.022 \), which was computed without
taking account of LLM effects \cite{Morf_NdA_SdS_02}. The intercept
of the straight line gives an estimate of the gap reduction due to
disorder, \( \Gamma (5/2)\approx 1.24 \)K, which is only slightly
less than the estimate for the intrinsic gap itself. We emphasize
that the estimate for \( \Gamma  \) for the two states is based solely
on the assumption that the \( 7/2 \) and \( 5/2 \) states are particle-hole
conjugates of each other and that the Coulomb interaction dominates
the value of the intrinsic gap. Although, in practice, the state at
\( \nu =7/2 \) is likely to be more strongly affected by LLM than
that at \( \nu =5/2 \), the assumption of particle-hole symmetry
between the two states should still be approximately valid. In the
following, we show that the effects of LLM reduce the theoretical
value to \( \delta ^{th}(5/2)\approx 0.016 \), so that the discrepancy
between theoretical and experimental estimates of the gap at 5/2 essentially
disappears. This provides further support for the identification of
the FQH state at \( \nu =5/2 \) as a paired state\cite{morf98, Rez-Hald00}.

\begin{figure}[t]
\resizebox*{6.3cm}{4.7cm}{\includegraphics{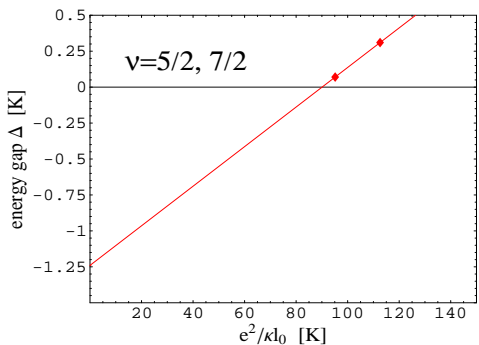}}

\caption{\label{fig25}The activation gaps, \protect\( \Delta ^{a}(\nu )\protect \)
from reference \cite{eisenstein_2002} plotted against \protect\( E_{c}=e^{2}/\kappa \ell _{0}\protect \)
for \protect\( \nu =5/2\protect \) (right) and \protect\( 7/2\protect \)
(left). The slope of the straight line through the measured gaps (cf.
eq. (\ref{eq:ansatz})) yields a coefficient \protect\( \delta (5/2)=0.014\protect \),
i.e. \protect\( \Delta _{i}=0.014\, E_{c}\protect \), and via the
intercept, an estimate of the gap reduction due to disorder \protect\( \Gamma _{5/2}^{est}\approx 1.2\protect \)K.}
\end{figure}

We have also reanalyzed older results for FQH states at filling \( \nu =p/(2p+1) \)
and \( (p+1)/(2p+1) \). We define the gap reduction \( \Gamma (\nu ) \),
as the difference between the measured activation gaps, \( \Delta ^{a}(\nu ) \),
and the intrinsic gap \( \Delta ^{i}(\nu )=\delta (\nu )E_{c} \):

\begin{equation}
\label{eq:definition}
\Delta ^{a}(\nu )=\delta (\nu )E_{c}-\Gamma (\nu ).
\end{equation}

As for the 5/2 system, we assume that the dominant contribution to
the FQH gaps comes from the Coulomb interaction of the electrons,
so that the symmetry related states in the set \( S_{\nu }=\{\nu ,\, (1-\nu ),\, 1+\nu ,\, 2-\nu \} \)
are all described by the same coefficient \( \delta _{\nu } \). If
in addition, we assume that the gap reduction \( \Gamma (\nu ') \)
has the same value \( \Gamma _{\nu } \) for all states \( \nu ' \)
in \( S_{\nu } \), the activation gap is approximated by

\begin{equation}
\label{eq:ansatz}
\Delta ^{a}(\nu ')\approx \delta _{\nu }E_{c}-\Gamma _{\nu }\, \, \, \, \, \, \, \, \, \, \, \forall \, \, \, \nu '\, \, \in \, \, S_{\nu }.
\end{equation}
 This allows us to use the measured transport gaps \( \Delta ^{a}(\nu ') \)
at \( \nu '=\nu ,\, (1-\nu ),... \) to obtain estimates, \( \delta ^{est}_{\nu } \)
and \( \Gamma ^{est}_{\nu } \), for the intrinsic gap and the gap
reduction for each family \( S_{\nu } \). 

The disorder scattering in the GaAs heterostructures, for which studies
of the activated transport have been reported, is thought to be due
mainly to ionized donors separated from the electron gas by a spacer
layer of width \( d \), where \( d\sim 800\AA  \) \cite{Du93}.
Our assumption that \( \Gamma _{\nu } \) is the same for all states
\( \nu ' \) in \( S_{\nu } \) is equivalent to assuming that, for
a given sample, the effect of these ionized donors on the FQH states
and the low-lying excitations will be the same for the FQH states
at filling fractions in the set \( S_{\nu } \). In the limit \( l_{0}/d\ll 1 \)
the response of the system on the length scale \( d \) should be
close to that of point-like quasiparticles, so that \( \Gamma _{\nu } \)
may depend on properties of the low-lying excitations like their charge
\( q \), which will be common to symmetry related states in the set
\( S_{\nu } \) but different for inequivalent sets \( S_{\nu } \).
However effects related to the internal structure of the excitations,
which would result in an additional dependence of \( \Gamma _{\nu } \)
on the ratio \( l_{0}/d \), are assumed to be small. When we compare
with our results from exact diagonalizations, this assumption appears
reasonable for the filling fractions close to 1/2 (see below).
\begin{figure}[t]
{\centering \resizebox*{7.2cm}{11.5cm}{\includegraphics{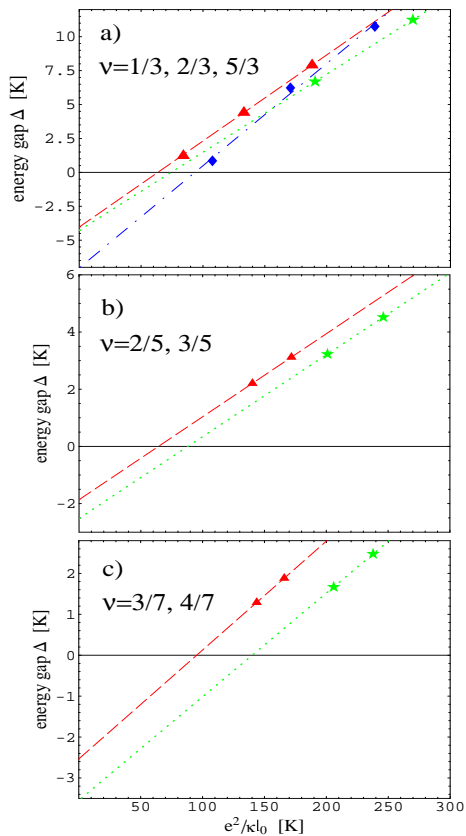}} \par}

\caption{\label{Fig1}The measured activation gaps, \protect\( \Delta ^{a}(\nu )\protect \)
plotted against the Coulomb energy \protect\( E_{c}=e^{2}/\kappa \ell _{0}.\protect \)
Triangles and asterisks respectively refer to samples A and B in \cite{Du93,Du_97},
diamonds represent results from \cite{Willett_88}. (a) Gaps at \protect\( \nu =1/3,\, 2/3,\, 5/3\protect \).
(b) Gaps at \protect\( \nu =2/5,\, 3/5.\protect \) (c) Gaps at \protect\( \nu =3/7,\, 4/7.\protect \)
The slope of \protect\( \Delta ^{a}\protect \) vs. \protect\( E_{c}\protect \)
gives an estimate of the intrinsic gap \protect\( \Delta ^{i}\approx \delta _{\nu }E_{c}\protect \)
of the set \protect\( S_{\nu }\protect \) of symmetry related states.
Samples of differing quality lead to similar slopes \protect\( \delta _{\nu }\protect \)
of the straight line fit, but to different intersections at \protect\( E_{c}=0\protect \),
which provide estimates for the gap reduction \protect\( \Gamma ^{est}_{\nu }\protect \)
for that family \protect\( S_{\nu }\protect \).}
\end{figure}

In Figure 2, we show the measured gaps taken from \cite{Du93, Du_97, Willett_88}
for three different very high-mobility samples at filling fractions
\( \nu =p/(2p+1),\, \nu '=1-\nu , \) and \( \nu '=2-\nu  \) for
\( p=1 \) (Figure 2a), \( p=2 \) (Figure 2b) and \( p=3 \) (Figure
2c) as function of \( E_{c} \). Samples A and B of \cite{Du93} have
an electron density \( n_{S} \) of (1.12 and 2.3)\( \times 10^{11}/ \)cm\( ^{2} \)
and mobilities \( \mu = \)(6.8 and 12)\( \times 10^{6} \)cm\( ^{2} \)/Vs.
The sample of \cite{Willett_88} (which we will call sample C) has
\( n_{S}=1.65\times 10^{11}/ \)cm\( ^{2} \) and a mobility \( \mu =5\times 10^{6} \)cm\( ^{2} \)/Vs.
Gaps at \( \nu =5/3 \) and \( 4/3 \) were reported for sample A
in \cite{Du_97}. The dependence of the gap on total magnetic field
in a tilted field experiment showed that at \( \nu =5/3 \) the ground
and low-lying excited states were spin-polarized. At \( \nu =4/3 \)
the ground state was not spin-polarized for tilt angles up to \( 65.1^{\circ } \)
while the excitations involved spin-reversals up to even larger tilt
angles. We therefore assume that only the states at \( \nu =1/3 \),
\( 2/3 \) and \( 5/3 \) are related by symmetry and not the state
at \( \nu =4/3 \). We show the measured gaps in untilted field as
a function of \( E_{c} \) in Figure \ref{Fig1}a. We note that the
slopes of the three straight line fits through the gaps of samples
A,B and C are quite similar yielding estimates of \( \delta _{1/3}= \)0.064,
0.058 and 0.075, respectively. By contrast, the intercepts at \( E_{c}=0 \),
which yield estimates, \( \Gamma ^{est}_{1/3} \), that vary by almost
a factor 2, and reflect differences in sample quality. In Figures
2b and 2c we show the analysis of the states states at \( \nu =2/5,\, 3/5 \)
and \( \nu =3/7,\, 4/7 \), respectively. We again note that the slope
of the gaps as function of \( E_{c} \) are very similar for the two
samples A and B. They yield estimates of \( \delta _{2/5}=0.029 \)
for both samples (Figure 2b) and \( \delta _{3/7}=0.027 \) and 0.025
(Figure 2c) for samples A and B, respectively.

An alternative approach to determining \( \Gamma (\nu ) \) is simply
to attribute the difference between precise calculations of FQH gaps
of disorder-free systems and measured gaps to the effects of disorder
and use equation \ref{eq:definition} as definition of \( \Gamma (\nu ) \).
Such calculations must of course include the effects of the finite
thickness \( w \) of the two-dimensional electron system as well
as LLM. We take account of LLM within the random phase approximation
for the dielectric function \cite{aleiner95}

\begin{equation}
\label{eq:dielectric}
\epsilon (q,\omega )=1-\tilde{V}(q)\, \Pi (q,\omega ),
\end{equation}
and represent the electron-electron interaction by

\begin{equation}
\label{eq:interaction}
U(r)=\int \frac{d^{2}q}{(2\pi )^{2}}\tilde{V}(q)\frac{e^{i\vec{q}\vec{r}}}{\epsilon (q,0)},
\end{equation}
where

\begin{equation}
\label{eq:vtwiddle}
\tilde{V}(q)=\frac{2\pi e^{2}}{\kappa q}e^{q^{2}w^{2}}erfc(qw)
\end{equation}
is the interaction between electrons that are trapped at the interface
in a Gaussian wave function of width \( w \) \cite{Morf_NdA_SdS_02}.
The polarization \( \Pi (q,\omega ) \) in (\ref{eq:dielectric})
is given by

\begin{eqnarray*}
\Pi (q,\omega ) & = & -\frac{m^{*}}{\pi \hbar ^{2}}\sum _{s}\, \, \sum ^{[\nu (s)-1]}_{n=0}F(\nu (s)-n)\, \sum ^{\infty }_{k=n+1}F(k-\nu (s))\\
 & \times  & \frac{(-1)^{(k-n)}(k-n)}{(\omega /\omega _{c})^{2}+(k-n)^{2}}L^{k-n}_{n}(x)\, L^{n-k}_{k}(x)\, e^{-x}\, \, \, \, \, \, \, \, \, \, \, \, \, \, \, \, \, \, \, \, (7)
\end{eqnarray*}
where \( x=(q\ell _{0})^{2}/2 \) and \( \sum _{s} \) stands for
the summation over spin \( s=\uparrow ,\downarrow  \). The symbol
\( [x] \) denotes the largest integer \( \leq x \). Equation (7)
agrees with expression (A1) of Aleiner and Glazman \cite{aleiner95},
which describes the spin degenerate case, \( \nu (\uparrow )=\nu (\downarrow )=N \),
with integer filling \( 2N \). The function \( F(z) \) is introduced
to treat the case of fractional filling and measures the filling fraction
of the Landau level \( n \), via \( F(z)=z \) for \( 0<z<1 \),
\( F(z)\equiv 1 \) for \( z\geq 1 \) and \( F(z)\equiv 0 \) for
\( z\leq 0 \). We have verified this method for incorporating finite
width and LLM corrections at filling fraction \( \nu =1/3 \) where
we could check that our results are consistent with those by Yoshioka
\cite{yoshioka_mixing}. Expression (\ref{eq:interaction}) together
with (\ref{eq:vtwiddle}) and (\ref{eq:dielectric}) lead to a modification
of the electron interaction at short separation, which is controlled
by the dimensionless parameter \( \lambda =E_{c}/\hbar \omega _{c} \).
Here \( \omega _{c}=eB/m^{*}c \) stands for the cyclotron frequency,
with \( m^{*} \) the effective mass of the electrons. A full account
of this method will be published elsewhere.

\begin{table}[t]
\begin{tabular}{|cc|c|l|cc|cc|}
\hline 
\( \nu  \)&
\( \nu ' \)&
Ref&
\( \delta _{\nu }^{est} \)&
\( \delta ^{th}(\nu ) \)&
\multicolumn{1}{c|}{\( \delta ^{th}(\nu ') \)}&
\multicolumn{1}{c}{\( \Gamma (\nu ) \)}&
\multicolumn{1}{c|}{\( \Gamma (\nu ') \)}\\
\hline 
1/3&
2/3&
\cite{Willett_88}&
0.069(9)&
0.75&
0.074&
7.2K&
6.4K\\
\hline 
1/3&
5/3&
\cite{Willett_88}&
0.077(4)&
0.075&
0.057&
7.2K&
5.3K\\
\hline
1/3&
2/3&
\cite{Du93}A&
0.063(2)&
0.077&
0.073&
6.5K&
5.3K\\
\hline
1/3&
5/3&
\cite{Du93,Du_97}A&
0.064(1)&
0.077&
0.052&
6.5K&
3.1K\\
\hline 
2/5&
3/5&
\cite{Du93}A&
0.029(3)&
0.036&
0.034&
3.0K&
2.6K\\
\hline 
3/7&
4/7&
\cite{Du93}A&
0.027(5)&
0.025&
0.025&
2.3K&
2.3K\\
\hline 
4/9&
5/9&
\cite{Du93}A&
0.013(6)&
0.019&
0.019&
2.2K&
2.1K\\
\hline
1/3&
2/3&
\cite{Du93}B&
0.058(2)&
0.077&
0.076&
9.4K&
7.8K\\
\hline 
2/5&
3/5&
\cite{Du93}B&
0.029(3)&
0.036&
0.036&
4.3K&
3.9K\\
\hline 
3/7&
4/7&
\cite{Du93}B&
0.025(3)&
0.025&
0.025&
3.6K&
3.5K\\
\hline 
4/9&
5/9&
\cite{Du93}B&
0.007(5)&
0.020&
0.020&
3.3K&
3.0K\\
\hline
5/2&
7/2&
\cite{eisenstein_2002}&
0.014&
0.016&
0.015&
1.5K&
1.4K\\
\hline
\end{tabular}

\caption{\label{Tab:Expt_vs_Theory}The values for the intrinsic gap \protect\( \delta ^{est}_{\nu }\protect \)
(see \ref{eq:deltai}) estimated by fitting the measured activation
gaps \protect\( \Delta ^{a}\protect \) to the Ansatz (\ref{eq:ansatz})
for different samples, together with our theoretical values \protect\( \delta ^{th}(\nu )\protect \)
for the gaps obtained from exact diagonalization studies (cf. \cite{Morf_NdA_SdS_02})
and the corresponding gap reduction \protect\( \Gamma (\nu )\protect \)
(see text). The theoretical values \protect\( \delta ^{th}(\nu )\protect \)
include finite width and Landau level mixing corrections. Numbers
in parentheses denote the error of the last quoted digit of \protect\( \delta _{\nu }^{est}\protect \).
These are calculated from the quoted experimental error \cite{Willett_88}
or from the discretization error when extracting numerical data from
experimental plots \cite{Du93,Du_97}. For the latter as well as for
\protect\( \nu =5/2\protect \) and 7/2, no experimental uncertainty
is specified. If for the values of \protect\( \Delta ^{a}\protect \)
of references \cite{Du93,Du_97} similar uncertainties are assumed
as specified in reference \cite{Willett_88}, errors for samples A
and B are 4-5 times bigger than quoted.}
\end{table}

For the particular sample of \cite{eisenstein_2002} we have repeated
the calculation of \cite{Morf_NdA_SdS_02} using the interaction amended
for LL mixing and taking account of the non-zero thickness of the
wavefunction. We have computed the quasiparticle and quasihole energies
as described in \cite{Morf_NdA_SdS_02} and, by extrapolating to the
thermodynamic limit, we have estimated the intrinsic gaps at \( \nu =5/2 \)
and \( \nu =7/2 \). The gaps are \( \delta ^{th}(5/2)=0.016 \),
and \( \delta ^{th}(7/2)=0.015 \), and are close to the estimate
\( \delta _{5/2}^{est}\approx 0.14 \) obtained from the experimental
values of \( \Delta ^{a} \) at \( \nu =5/2 \) and 7/2, using (\ref{eq:ansatz})
as discussed above. We have also calculated the finite width and LLM
corrections for all the other states reported in references \cite{Willett_88,Du93,Du_97}.
These results are listed in Table 1. As can be seen, the estimates
\( \delta _{\nu }^{est} \) of the gap coefficients, calculated on
the basis of the simple Ansatz (\ref{eq:ansatz}) from the experimental
values of the activation gap \( \Delta ^{a} \), are consistent with
the theoretical values \( \delta ^{th}(\nu ) \) within realistic
error bars (cf. caption of Table \ref{Tab:Expt_vs_Theory}). We also
note that the gap coefficients \( \delta ^{th}(\nu ) \) for symmetry
related states are indeed very similar even when finite width and
LLM effects are taken into account. With the exception of the \( \nu =5/3 \)
case, our assumption \( \delta (\nu ')\approx \delta (\nu ) \) for
all \( \nu  \) in the set \( S_{\nu } \) of symmetry related states
therefore appears reasonable.

\begin{figure}[t]
{\centering \resizebox*{6cm}{!}{\includegraphics{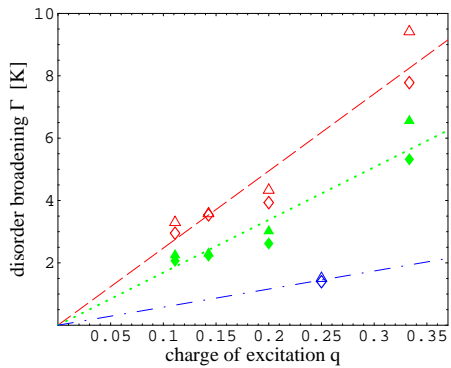}} \par}

\caption{\label{Fig:Gamma_vs_Charge}\protect\( \Gamma (\nu )\protect \)
for various \emph{}samples of differerent high electron mobility plotted
as a function of the charge of the elementary excitations. The full
symbols refer to sample A, open triangles and diamonds to sample B
\cite{Du93,Du_97}, the data on the dash-dotted line represent \protect\( \nu =5/2\protect \)
and 7/2 \cite{eisenstein_2002}. The triangles refer to \protect\( \nu ,\protect \)
diamonds to \protect\( 1-\nu \protect \).}
\end{figure}

Estimates for \( \Gamma (\nu ) \), based on equation (\ref{eq:definition}),
are listed in the last two columns of Table \ref{Tab:Expt_vs_Theory}
and plotted in Figure \ref{Fig:Gamma_vs_Charge} against the charge
\( q=1/(2p+1) \) of the elementary excitation of the FQH state at
\( \nu =p/(2p+1) \) or its symmetry related siblings. The values
we obtain for \( \Gamma (\nu ) \) scale with the charge of the excitation
in each sample. They are comparable for families of symmetry related
states as we assumed in (\ref{eq:ansatz}), with the assumption better
justified the closer the filling fractions are to \( \nu =1/2 \).
However, in each family, \( \Gamma (\nu ) \) is systematically smaller
for larger filling factors, with smaller differences between \( \Gamma (\nu ) \)
and \( \Gamma (1-\nu ) \) when \( \nu  \) is close to 1/2. The largest
differences are for the case of \( \nu =1/3,\, 2/3 \) and 5/3, where
\( \Gamma (1/3) \) is about twice \( \Gamma (5/3) \). We attribute
this reduction of \( \Gamma (\nu ) \) at larger \( \nu  \) to LLM
and other polarization/screening effects which should increase as
\( l_{0}/d \) increases (for the \( \nu =5/3 \) state in \cite{Du_97}
\( l_{0}/d\sim 1/3 \)), but which would require a microscopic model
of the response of the FQH system to potential variations to quantify. 

A reduction of the gap as a result of disorder is reminiscent of what
happens in models of the effects of disorder on the integer quantum
Hall effect (IQHE) in the variable range hopping regime \cite{Polyakov_Shklovskii_94/5}.
However, this model is not thought to apply in the regime of large
\( d \), and may not be adequate to describe the gap reductions that
have been observed.

The authors acknowledge useful discussions with B.I. Halperin.

\end{document}